\documentclass[%
 reprint,
 amsmath,amssymb,
 aps,
 prl,
]{revtex4-2}

\usepackage{graphicx}
\usepackage{dcolumn}
\usepackage{bm}
\usepackage{siunitx}
\DeclareSIUnit{\molar}{M}
\usepackage[hidelinks]{hyperref}

\begin{document}

\title{Observation of Dynamic Nuclear Polarization Echoes}

\author{Nino Wili}
 \email{science@ninowili.ch}
\author{Anders B. Nielsen}
\author{José P. Carvalho}
\author{Niels Chr. Nielsen}
 \email{ncn@chem.au.dk}
\affiliation{%
Interdisciplinary Nanoscience Center (iNANO) and Department of Chemistry, Aarhus University, Gustav Wieds Vej 14, DK-8000 Aarhus C, Denmark
}%

\date{\today}

\begin{abstract}
It is demonstrated that the time evolution of the electron-nuclear polarization transfer process during pulsed dynamic nuclear polarization (DNP) can be reversed on a microsecond timescale, leading to the observation of DNP echoes. The DNP echoes are induced by consecutive application of two pulse trains that produce effective Hamiltonians that differ only in the sign of the effective hyperfine coupling. The experiments have been performed on a frozen solution of trityl radicals in water/glycerol on a home-built X-band EPR/DNP spectrometer at \SI{80}{K}. We envisage that  DNP echoes will play an important role in future development of pulsed DNP for sensitivity-enhanced NMR, hyperfine spectroscopy, and quantum sensing.

\end{abstract}

\maketitle


\section{Introduction}
Echo phenomena are a cornerstone of coherent spectroscopy. They are central to magnetic resonance, including nuclear magnetic resonance (NMR) \cite{abragamPrinciplesNuclearMagnetism1961,ernstPrinciplesNuclearMagnetic1987} and electron paramagnetic resonance (EPR) \cite{pontiEchoPhenomenaElectron1994,schweigerPrinciplesPulseElectron2001}, but echoes have also been observed in Ruby lasers \cite{kurnitObservationPhotonEcho1964a} and in microwave \cite{glorieuxMicrowaveEchoesInhomogeneous1976} and infrared spectroscopy \cite{bermanCoherentOpticalTransient1975a}. In magnetic resonance, echoes are formed by \textit{refocusing} an interaction in the spin Hamiltonian that leads to an \textit{apparent} signal decay. For example, the pioneering Hahn echo \cite{hahnSpinEchoes1950} refocuses heteronuclear spin-spin couplings and inhomogeneities in the external field. Another important example is the magic echo and variants thereof, which have been shown to invert the evolution in a homonuclear system of strongly coupled spins \cite{schneiderNegativeTimeDevelopment1969,rhimViolationSpinTemperatureHypothesis1970}. Remarkably, this observation violates the spin temperature hypothesis and highlights that many ``relaxation'' phenomena in magnetic resonance are in fact coherent in nature and accordingly may be dealt with through pulsed operation, and refocused at least in principle. This phenomenon has been referred to as ``time-reversal'' experiments \cite{rhimTimeReversalExperimentsDipolarCoupled1971a}, as the system evolves as if time was reversed.

Besides spin echoes, also polarization and coherence transfer elements are fundamental both in NMR and EPR \cite{ernstPrinciplesNuclearMagnetic1987,schweigerPrinciplesPulseElectron2001}. In NMR of solids, the most prominent polarization transfer scheme is cross-polarization (CP) \cite{pinesProtonEnhancedNuclear1972}, where two dipolar coupled heteronuclei are spin locked by radio-frequency (rf) fields with the same nutation frequency, known as Hartmann-Hahn matching \cite{hartmannNuclearDoubleResonance1962}. In the context of echoes, it has been shown  that it is possible to invert the sign of the effective heteronuclear coupling Hamiltonian by strategically choosing the phases and offsets of both spin locks, leading to so-called cross-polarization echoes \cite{ernstTimereversalCrosspolarizationNuclear1998}.

In spin systems containing electron spins and nuclear spins, the electron-nuclear polarization transfer is known as dynamic nuclear polarization (DNP) \cite{abragamPrinciplesDynamicNuclear1978}. Historically, most DNP experiments have been conducted with continuous wave (CW) microwave (mw) irradiation. This is especially the case for high-field, high-resolution DNP-enhanced NMR aimed at investigating chemical and biological systems with orders of magnitude higher sensitivity than reachable by conventional NMR \cite{malyDynamicNuclearPolarization2008,lillythankamonyDynamicNuclearPolarization2017}. 
\textit{Pulsed} DNP was introduced in 1987 with the Nuclear spin Orientation via Electron spin Locking (NOVEL) pulse sequence \cite{brunnerCrossPolarizationElectron1987,henstraNuclearSpinOrientation1988}. This experiment is the analogon of the CP experiment, but because of the large difference between gyromagnetic ratios of electron and nuclear spins, polarization transfer through the pseudosecular hyperfine coupling is in this case driven by matching the spin lock Rabi frequency of the electrons to the Larmor frequency of the nuclei. In this particular case no irradiation of the nuclei is required. In recent years, many other pulsed DNP schemes have been developed \cite{schwartzRobustOpticalPolarization2018,tanTimeoptimizedPulsedDynamic2019,redrouthuEfficientPulsedDynamic2022,wiliDesigningBroadbandPulsed2022,redrouthuDynamicNuclearPolarization2023}, and it was recognized that the description of pulsed DNP sequences in many respects resembles that of magic-angle spinning (MAS) recoupling sequences \cite{tanTimeoptimizedPulsedDynamic2019}. In the latter case, the pulse sequences interfere with the physical sample rotation. In pulsed DNP, the pulse sequence interferes with the rotation in spin space due to the nuclear Zeeman interaction.
 
 In this work, we show that it is possible to invert the effective hyperfine couplings in pulsed DNP, which leads to the formation of dynamic nuclear polarization echoes. Such echoes may have important applications in further development of pulsed DNP for atomic structure analysis and emerging quantum technologies.
 
 \section{Overall idea}
 We consider a system of one electron spin and $N_I$ nuclear spins of the same isotope.  The Hamiltonian of this system, in the electron spin rotating frame and employing the high-field approximation for the electron spin, is given in angular frequencies by
 \begin{align}
 	\mathcal{H}&=\Delta\omega_S S_z + \mathcal{H}_\text{mw} \nonumber\\
 	&+\sum_{i}^{N_I} \omega_I I_{iz} + S_z\left(A_x^{i} I_{ix}+A_y^{i} I_{iy}+A_z^{i} I_{iz}  \right) \nonumber \\
 		&+\mathcal{H}_\text{nn} ,
 \end{align}
where $\Delta\omega_S$ is the electron spin offset, $\mathcal{H}_\text{mw}$ describes the microwave irradiation, $\omega_I=-\gamma_I B_0$ is the nuclear Zeeman frequency, $A^i$ is the anisotropic hyperfine interaction between the electron and nucleus $i$, and $\mathcal{H}_\text{nn}$ corresponds to the nuclear-nuclear couplings.

Our goal is to generate an effective Hamiltonian that leads to electron-nuclear spin polarization transfer, and then to invert the sign of the effective Hamiltonian to create echoes which can be observed on either of the two spin species. This should be achieved by irradiating the electrons only. We will restrict ourselves to periodic irradiation schemes, where $\mathcal{H}_\text{mw}(t+\tau_\text{m})=\mathcal{H}_\text{mw}(t)$ with $\tau_\text{m}$ being the period of the Hamiltonian. It was shown by single-spin vector analysis and average Hamiltonian theory\cite{shankarGeneralTheoreticalDescription2017,nielsenSinglespinVectorAnalysis2019,wiliDesigningBroadbandPulsed2022,nielsenAccurateAnalysisPerspectives2022} and operator-based Floquet theory \cite{tanTimeoptimizedPulsedDynamic2019}, that the first-order effective Hamiltonian governing the polarization transfer in DNP can be written as
 \begin{align}
	\mathcal{H}_\text{eff}&=-\omega_\text{eff}^{(S)} \tilde{S}_z + \omega_\text{eff}^{(I)} I_z \nonumber\\
	&+\sum_{i}^{N_I} \frac{B_i}{4} \left( \left( a_{\Delta}  \tilde{S}^+I^- + a_{\Delta}^*\tilde{S}^-I^+ \right) \right. \nonumber\\
	&+ \left.
	 \left( a_{\Sigma}\tilde{S}^+I^+ + a_{\Sigma}^*\tilde{S}^-I^-\right)\right) \nonumber\\
	&+\tilde{\mathcal{H}}_\text{nn} ,
\end{align}
with $B=\sqrt{A_x+A_y}$. The tilde indicates an \textit{effective} frame, where the $z$-axis points along the overall axis of rotation over one period of the periodic pulse sequence. The magnitude of the effective fields is given by $\omega_\text{eff}=\beta_\text{eff}/\tau_\text{m}$, where $\beta_\text{eff}$ is the overall angle of rotation. We choose the convention $|\beta_\text{eff}|\leq\pi$, such that the effective field is at most half the modulation frequency, $|\omega_\text{eff}|\leq\ \omega_\text{m}/2=\pi/\tau_\text{m}$. If the nuclear spins are not irradiated, then the nuclear effective field is $\omega_\text{eff}^{(I)}=\omega_I-\text{round}\left(\omega_I/\omega_\text{m}\right)\omega_\text{m}$. The strength of the effective hyperfine coupling is encoded in the scaling factors $a_\Delta$ and $a_\Sigma$, which depend on the details of the electron spin trajectory and are complex in general \cite{wiliDesigningBroadbandPulsed2022}. 

Depending on the size and signs of the electron or nuclear spin effective fields, the polarization can be transferred in the zero-quantum (ZQ), or double-quantum (DQ) subspaces. Usually it is desirable that only one of the two takes place, which is the case either if $|\omega_\text{eff}^{(S)}|\approx|\omega_\text{eff}^{(I)}|\gg |B_i|$ or if $|\omega_\text{eff}^{(S)}|\approx|\omega_\text{eff}^{(I)}|\approx 0$  and either $a_\Delta=0$ or $a_\Sigma=0$ \footnote{Another special case is $|\omega_\text{eff}^{(S)}|\approx|\omega_\text{eff}^{(I)}|\approx \omega_\text{m}/2$. In this case there are additional time-independent terms in the interaction frame that have to be considered. However, in this case one could look at two periods of the repeating element, which would again lead to $|\omega_\text{eff}^{(S)}|\approx|\omega_\text{eff}^{(I)}|\approx 0$.}.

Let us assume that we choose a ZQ-resonance condition with a sizable effective field and a negative Zeeman frequency of the nuclei. In this case, the DQ terms in the effective Hamiltonian are truncated, and we are left with 
 \begin{align}
	\mathcal{H}_\text{eff}&=-\omega_\text{eff}^{(S)} \tilde{S}_z - |\omega_\text{eff}^{(I)}| I_z \nonumber\\
	&+\sum_{i}^{N_I} \frac{B_i}{4} \left( a_{\Delta}  \tilde{S}^+I^- + a_{\Delta}^*\tilde{S}^-I^+ \right) \nonumber \\
	&+\tilde{\mathcal{H}}_\text{nn} .
\end{align}
So far, the action of this type of Hamiltonians has essentially only been considered for the $\tilde{S}_z \rightarrow I_z$ transfer corresponding to an inversion ($\pi$ rotation) in ZQ space. Accordingly, if only these two operators are of interest, then the phase of $a_\Delta$ does not matter. However, for our goal, this phase is essential, since it allows for an inversion of the coupling Hamiltonian. \textit{If} the effective field of the electron spins points along $z$, the phase of $a_\Delta$ can simply be controlled by the overall phase of the mw pulse sequence. This phase change corresponds to a $S_z$ rotation, and a phase change of $\pi$ leads to
\begin{align}
& e^{i\pi S_z}\sum_{i}^{N_I} \frac{B_i}{4} \left( a_{\Delta}  {S}^+I^- + a_{\Delta}^*{S}^-I^+ \right)e^{-i\pi S_z}\nonumber\\
     &=- \sum_{i}^{N_I} \frac{B_i}{4} \left( a_{\Delta}  {S}^+I^- + a_{\Delta}^*{S}^-I^+ \right) .
\end{align}

The effective fields and the nuclear coupling Hamiltonian are not affected by this phase change. However, the DNP echo formation still occurs if 1) the mismatch of the effective field is not too large, and 2) the timescale of the electron-nuclear transfer is much shorter than any nuclear-nuclear spin dynamics. Condition 1) should be mostly fulfilled for reasonable pulse sequences and narrow EPR spectra, and condition 2) may be fulfilled by choosing reasonably short DNP echo times. We should note that the approach of inverting the coupling Hamiltonian is fundamentally different from inverting the effective field. The latter approach changes the sequence from a ZQ to a DQ one. The dynamics then occur in different subspaces, and no echo would be formed. For example, simply inverting the phase of the spin lock in NOVEL does \textit{not} lead to the formation of a DNP echo.

\section{DNP pulse sequence}
\begin{figure}
	\includegraphics[width=\linewidth]{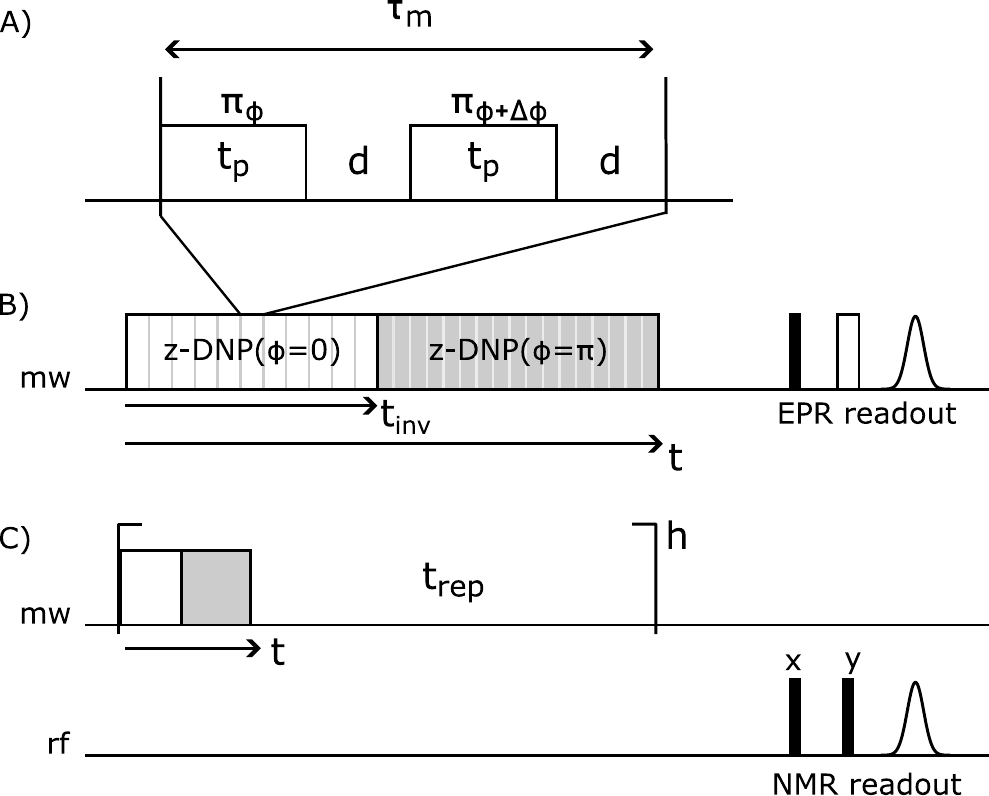}
	\caption{\label{fig:sequences} Pulse sequences used in this work. A) Basic DNP building block. Two $\pi$ pulses of length $t_p$ with a phase difference of $\Delta\phi$, separated by a delay $d$. The overall length of one element is $\tau_m=2(t_p+d)$. The sequence leads to an overall rotation around $S_z$ by an angle of $2\Delta\phi$. B) Generation of DNP echo and EPR readout. The basic building block in A) is repeated for a time $t_\text{inv}$, at which point the overall phase is inverted. The sequence is then continued up to a total time $t$.  At this point, the electron polarization is detected via a Hahn echo.  The DNP echo forms for $t=2t_\text{inv}$. C) Formation of DNP echo and detection on the nuclear spins. The sequence in B), excluding the EPR readout, is repeated $h$ times, with a repetition time $t_\text{rep}$ on the the order of the electron $T_{1,e}$. The polarization of the bulk protons is then detected with a solid echo. The residual bulk proton polarization is destroyed by a saturation train after each readout. Filled and open pulses in  readout corresponds to $\pi/2$ and $\pi$ pulses, respectively.}
\end{figure}

In principle, any periodic DNP pulse sequence that acts only on the electron spins and generates an effective rotation around $z$ can be used. We chose a conceptually simple building block consisting of two $\pi$ pulses that differ in their phases by $\Delta\phi$, as illustrated in Fig.~\ref{fig:sequences} A). It can be looked at as a special case of TPPM DNP \cite{redrouthuDynamicNuclearPolarization2023} with added delays. The delay $d$ between the pulses is introduced to make the sequences less susceptible to pulse transients. Overall, the modulation time of the sequence is given by $\tau_\text{m}=2(t_p+d)$. During that time, the nuclear spins rotate around $I_z$ by an angle $\beta_\text{eff}^{(I)}=\omega_I\cdot2(t_p+d)$. For an on-resonance spin packet, the overall electron spin rotation is given by $\beta_\text{eff}^{(S)}=2\Delta\phi$. The two angles have to match up (modulo $2\pi$). We chose $\Delta\phi=(2\pi-\omega_I\tau_\text{m})/2$. More concretely, with a nuclear Zeeman frequency of $\omega_I/2\pi=$\SI{-14.787}{\mega\hertz}, we used $t_p=$\SI{12}{\nano\second}, $d=$\SI{10}{\nano\second}, and $\Delta\phi=1.098\approx$ \SI{63}{\degree}. The exact optimal value for $\Delta\phi$ was set experimentally, and could deviate from this prediction by a few degrees. The scaling factor for these values were numerically calculated to be $|a_\Delta|=0.3371$.

\section{Experimental}

All experiments were conducted on a home-built X-band pulsed EPR/DNP spectrometer (based on the design of Doll \textit{et al.} \cite{dollWidebandFrequencysweptExcitation2017}) with a sample of \SI{5}{\milli\molar} trityl (OX063) in a H$_2$O:D$_2$O:Glycerol-d$_5$ solution (1:3:6 by volume) at 80 K. This is a common system for pulsed DNP studies \cite{mathiesPulsedDynamicNuclear2016}, although usually with a slightly lower degree of solvent protonation. A commercial MD4 electron-nuclear double resonance probe (Bruker BioSpin) extended with an external tuning and matching circuit was used, and NMR signals were detected with a Spincore iSpin-NMR console (SpinCore Technologies, Inc., Gainesville, FL).

\section{Results}

\begin{figure*}[!]
	\includegraphics[width=\linewidth]{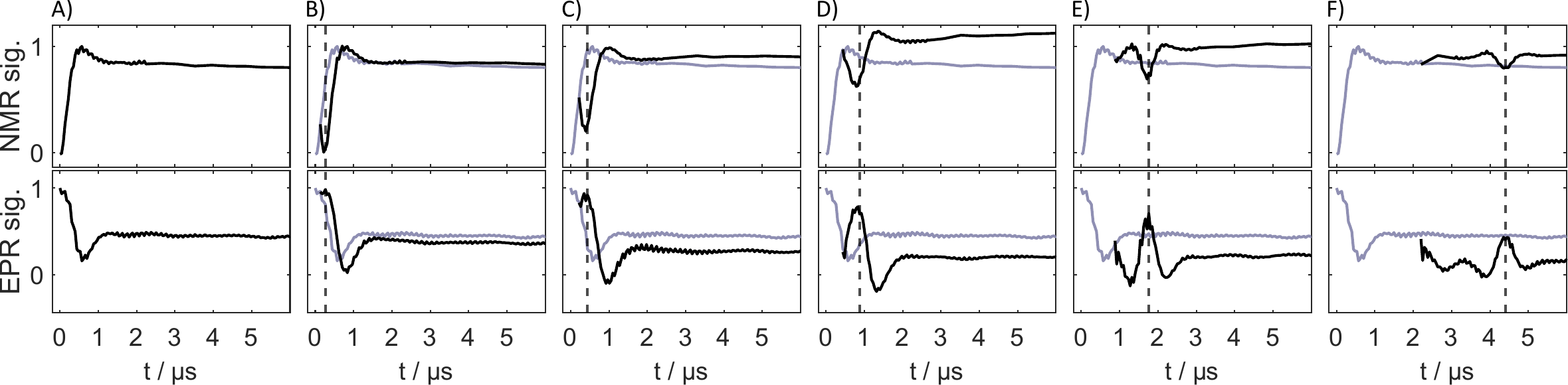}
	\caption{\label{fig:dnp_echoes} Observation of DNP echoes via NMR and EPR read out. \textit{Top}: NMR signals \textit{Bottom}: Corresponding EPR signals. A) Reference signals for $t_\text{inv}=0$. B)-F) $t_\text{inv}=$\SI{132}{\nano\second}, \SI{220}{\nano\second}, \SI{440}{\nano\second}, \SI{880}{\nano\second}, and \SI{2200}{\nano\second}. The expected echo position at $t=2t_\text{inv}$ is marked by a dashed line. The reference signal from A) is shown in gray in the other panels. The apparent stop of small oscillations in the NMR signals after around \SI{2}{\micro\second} is due to a coarser sampling interval at longer times. }
\end{figure*}

 DNP echo formation for trityl was observed both on the electron and the nuclear spins. For the electron spins, we used the sequence shown in Fig.~\ref{fig:sequences} B). The basic building block described above was repeated for a time $t_\text{inv}$, at which point the overall phase of the building block was changed by $\pi$. After a total time $t$, and a waiting time of \SI{10}{\micro\second}, which is long enough for all remaining electron spin coherences to decay, but much shorter than the electron $T_{1,e}$ (about \SI{2}{\milli\second}), the electron polarization was read out with a simple Hahn echo sequence. 

For the nuclear polarization, it is not really possible to detect the protons that are very close to the unpaired electron. This is due to the  low concentration of unpaired electrons, fast nuclear spin relaxation, and hyperfine couplings that exceed the detection bandwidth. However, the electron-nuclear spin dynamics can be imprinted onto the bulk proton magnetization by making use of nuclear spin diffusion. For this to occur, one has to repeat the DNP pulse sequence a large number of times, see Fig.~\ref{fig:sequences} C) pulse sequence. The mw pulse sequence in Fig.~\ref{fig:sequences} B), excluding the Hahn echo, was repeated 1000 times with a repetition time of \SI{2}{\milli\second}. 
During the waiting times, the nuclear spin polarization diffuses to the bulk protons. This polarization is then read out on the nuclear spins by a solid-echo sequence.

The NMR and EPR signals observed for different phase inversion times $t_\text{inv}$ are shown in Fig.~\ref{fig:dnp_echoes}. Panel A) shows the polarization dynamics in the absence of the overall phase change, while panels B) - F) were recorded with increasing values for $t_\text{inv}$. One clearly observes the formation of DNP echoes at a time $2t_\text{inv}$, indicated by dashed lines. For short $t_\text{inv}$ times, \textit{i.e.} \SI{132}{\nano\second}, the inversion is essentially perfect, and the nuclear polarization goes back to zero at the expected echo position, while the the electron polarization goes back to the initial intensity. At increasingly long $t_\text{inv}$ times, the refocusing is no longer perfect, because the matching condition is not perfectly fulfilled for all spin packets. Additionally, higher-order terms of the effective coupling Hamiltonian can also lead to non-perfect refocusing. While the nuclear-nuclear spin couplings might also play a role, we expect this effect to be  negligible on the presented time scale. Notwithstanding, the echo formation is still clearly visible in all cases shown here.

One curiosity is that in some cases, the overall nuclear polarization obtained with the phase inversion is higher than without it, most prominently in Fig.~\ref{fig:dnp_echoes} D). While this was not the goal of this work, it does give some interesting hints if one want to obtain as high a nuclear polarization as possible in DNP experiments. These features are qualitatively reproduced in numerical simulations (see Supplementary Information). In the presence of mismatches due to electron offsets and mw inhomogeneity over the sample volume, the phase inversion can partially correct some imperfections, as it essentially acts as a composite pulse in the ZQ/DQ subspace.  

\section{Discussion and Outlook}
We have shown that it is possible to generate and observe dynamic nuclear polarization echoes. This is achieved by choosing a pulsed DNP sequence that forms an effective $z$-rotation of the electron spin. Inverting the overall phase of this sequence causes inversion of the electron-nuclear spin coupling term in the effective Hamiltonian responsible for the polarization transfer. This effectively lets the electron-nuclear spin system evolve backwards in time. The DNP echo formation can be observed both on the electrons and the nuclei. To supplement the observations in the preceding sections, we here add different aspects that may put the results into perspective and suggest avenues to further investigations and applications.

In principle, any pulse sequence leading to an overall $z$-rotation could be employed. The recently introduced PulsePol sequence \cite{schwartzRobustOpticalPolarization2018} should be a good alternative candidate. We did not pursue this in this work, because in our setup, the obtainable electron spin Rabi frequency is only about a factor of 2-3 larger than the nuclear Zeeman frequency. This implies that either the pulses would fill a significant amount of the modulation time, or the modulation time would get very long and thereby the scaling factor low. In the former case, the PulsePol sequence generates both ZQ and DQ transfer. In typical setups that investigate nitrogen-vacancy centers, the electron spin Rabi frequency is much larger than the nuclear Zeeman frequency of $^{13}$C, which alleviates this problem.

It might just as well be possible to observe DNP echoes with pulse sequences where the effective field does not point along $z$. In this case, however, the inversion of the coupling Hamiltonian is more complicated, since one has to change the phase of the pulse sequence in the effective frame, in a time that it short compared to the spin dynamics of the system. In the case of NOVEL, an infinitely short $\pi$ pulse along the spin-lock axis would work in theory. In practice, the spin dynamics during the pulse would need to be compensated somehow.

The pulse scheme proposed here inverts the effective hyperfine coupling during DNP, but it leaves the nuclear-nuclear couplings unaffected. In principle, it might be possible to invert the sign of the complete Hamiltonian, although we expect this to be rather challenging to realize in practice.

Serendipitously, we observed that a larger overall nuclear polarization can be achieved with the phase inversion. This observation could be explored further, \textit{i.e.} by employing known composite pulse schemes in the ZQ/DQ subspaces. 

We expect that the DNP echo could form the basis for pulse sequences that are used to detect nuclear spins in the vicinity of an electron spin. In the magnetic resonance community, this is known as hyperfine spectroscopy. Since the DNP echo and its decay depend on all hyperfine and nuclear-nuclear spin-spin couplings, it might also be possible to infer information about the coupling network around electron spins. Of course, it should also be possible to use DNP echoes in other quantum sensing approaches \cite{degenQuantumSensing2017}, including single-spin readout experiments such as the ones employed in nitrogen-vacancy centers \cite{gruberScanningConfocalOptical1997}.

\begin{acknowledgments}
\section{Acknowledgements}

\paragraph{Funding:}
	NW was supported by the Swiss National Science Foundation (Postdoc.Mobility grant 206623). We acknowlege support from the Novo Nordisk Foundation (NERD grant NNF22OC0076002) and the Aarhus University Research Foundation (AUFF, grant AUFF-E-2021-9-22).
	
\paragraph{Author contributions:}
All authors developed and discussed the theory. N.W. built the spectrometer, performed the measurements and wrote the initial draft. All authors helped reviewing the manuscript. 

\paragraph{Competing Interests:}
The authors declare that they have no competing interests.

\paragraph{Data and materials availability:}
All data needed to evaluate the conclusions in the paper are present in the paper and/or the Supplementary Materials. Raw data and processing scripts will be deposited on Zenodo upon acceptance of the manuscript.
\end{acknowledgments}


\bibliography{dnp_echo}

\begin{thebibliography}{32}%
\makeatletter
\providecommand \@ifxundefined [1]{%
 \@ifx{#1\undefined}
}%
\providecommand \@ifnum [1]{%
 \ifnum #1\expandafter \@firstoftwo
 \else \expandafter \@secondoftwo
 \fi
}%
\providecommand \@ifx [1]{%
 \ifx #1\expandafter \@firstoftwo
 \else \expandafter \@secondoftwo
 \fi
}%
\providecommand \natexlab [1]{#1}%
\providecommand \enquote  [1]{``#1''}%
\providecommand \bibnamefont  [1]{#1}%
\providecommand \bibfnamefont [1]{#1}%
\providecommand \citenamefont [1]{#1}%
\providecommand \href@noop [0]{\@secondoftwo}%
\providecommand \href [0]{\begingroup \@sanitize@url \@href}%
\providecommand \@href[1]{\@@startlink{#1}\@@href}%
\providecommand \@@href[1]{\endgroup#1\@@endlink}%
\providecommand \@sanitize@url [0]{\catcode `\\12\catcode `\$12\catcode
  `\&12\catcode `\#12\catcode `\^12\catcode `\_12\catcode `\%12\relax}%
\providecommand \@@startlink[1]{}%
\providecommand \@@endlink[0]{}%
\providecommand \url  [0]{\begingroup\@sanitize@url \@url }%
\providecommand \@url [1]{\endgroup\@href {#1}{\urlprefix }}%
\providecommand \urlprefix  [0]{URL }%
\providecommand \Eprint [0]{\href }%
\providecommand \doibase [0]{https://doi.org/}%
\providecommand \selectlanguage [0]{\@gobble}%
\providecommand \bibinfo  [0]{\@secondoftwo}%
\providecommand \bibfield  [0]{\@secondoftwo}%
\providecommand \translation [1]{[#1]}%
\providecommand \BibitemOpen [0]{}%
\providecommand \bibitemStop [0]{}%
\providecommand \bibitemNoStop [0]{.\EOS\space}%
\providecommand \EOS [0]{\spacefactor3000\relax}%
\providecommand \BibitemShut  [1]{\csname bibitem#1\endcsname}%
\let\auto@bib@innerbib\@empty
\bibitem [{\citenamefont
  {Abragam}(1961)}]{abragamPrinciplesNuclearMagnetism1961}%
  \BibitemOpen
  \bibfield  {author} {\bibinfo {author} {\bibfnamefont {A.}~\bibnamefont
  {Abragam}},\ }\href@noop {} {\emph {\bibinfo {title} {The {{Principles}} of
  {{Nuclear Magnetism}}}}},\ International {{Series}} of {{Monographs}} on
  {{Physics}}\ (\bibinfo  {publisher} {Oxford University Press},\ \bibinfo
  {address} {Oxford, New York},\ \bibinfo {year} {1961})\BibitemShut {NoStop}%
\bibitem [{\citenamefont {Ernst}\ \emph {et~al.}(1987)\citenamefont {Ernst},
  \citenamefont {Bodenhausen},\ and\ \citenamefont
  {Wokaun}}]{ernstPrinciplesNuclearMagnetic1987}%
  \BibitemOpen
  \bibfield  {author} {\bibinfo {author} {\bibfnamefont {R.~R.}\ \bibnamefont
  {Ernst}}, \bibinfo {author} {\bibfnamefont {G.}~\bibnamefont {Bodenhausen}},\
  and\ \bibinfo {author} {\bibfnamefont {A.}~\bibnamefont {Wokaun}},\ }\href
  {https://doi.org/10.1093/oso/9780198556473.001.0001} {\emph {\bibinfo {title}
  {Principles of {{Nuclear Magnetic Resonance}} in {{One}} and {{Two
  Dimensions}}}}}\ (\bibinfo  {publisher} {Oxford University Press},\ \bibinfo
  {year} {1987})\BibitemShut {NoStop}%
\bibitem [{\citenamefont {Ponti}\ and\ \citenamefont
  {Schweiger}(1994)}]{pontiEchoPhenomenaElectron1994}%
  \BibitemOpen
  \bibfield  {author} {\bibinfo {author} {\bibfnamefont {A.}~\bibnamefont
  {Ponti}}\ and\ \bibinfo {author} {\bibfnamefont {A.}~\bibnamefont
  {Schweiger}},\ }\bibfield  {title} {\bibinfo {title} {Echo phenomena in
  electron paramagnetic resonance spectroscopy},\ }\href
  {https://doi.org/10.1007/BF03162620} {\bibfield  {journal} {\bibinfo
  {journal} {Applied Magnetic Resonance}\ }\textbf {\bibinfo {volume} {7}},\
  \bibinfo {pages} {363} (\bibinfo {year} {1994})}\BibitemShut {NoStop}%
\bibitem [{\citenamefont {Schweiger}\ and\ \citenamefont
  {Jeschke}(2001)}]{schweigerPrinciplesPulseElectron2001}%
  \BibitemOpen
  \bibfield  {author} {\bibinfo {author} {\bibfnamefont {A.}~\bibnamefont
  {Schweiger}}\ and\ \bibinfo {author} {\bibfnamefont {G.}~\bibnamefont
  {Jeschke}},\ }\href@noop {} {\emph {\bibinfo {title} {Principles of Pulse
  Electron Paramagnetic Resonance}}}\ (\bibinfo  {publisher} {Oxford University
  Press},\ \bibinfo {address} {Oxford, UK ; New York},\ \bibinfo {year}
  {2001})\BibitemShut {NoStop}%
\bibitem [{\citenamefont {Kurnit}\ \emph {et~al.}(1964)\citenamefont {Kurnit},
  \citenamefont {Abella},\ and\ \citenamefont
  {Hartmann}}]{kurnitObservationPhotonEcho1964a}%
  \BibitemOpen
  \bibfield  {author} {\bibinfo {author} {\bibfnamefont {N.~A.}\ \bibnamefont
  {Kurnit}}, \bibinfo {author} {\bibfnamefont {I.~D.}\ \bibnamefont {Abella}},\
  and\ \bibinfo {author} {\bibfnamefont {S.~R.}\ \bibnamefont {Hartmann}},\
  }\bibfield  {title} {\bibinfo {title} {Observation of a {{Photon Echo}}},\
  }\href {https://doi.org/10.1103/PhysRevLett.13.567} {\bibfield  {journal}
  {\bibinfo  {journal} {Physical Review Letters}\ }\textbf {\bibinfo {volume}
  {13}},\ \bibinfo {pages} {567} (\bibinfo {year} {1964})}\BibitemShut
  {NoStop}%
\bibitem [{\citenamefont {Glorieux}\ \emph {et~al.}(1976)\citenamefont
  {Glorieux}, \citenamefont {Legrand},\ and\ \citenamefont
  {Macke}}]{glorieuxMicrowaveEchoesInhomogeneous1976}%
  \BibitemOpen
  \bibfield  {author} {\bibinfo {author} {\bibfnamefont {P.}~\bibnamefont
  {Glorieux}}, \bibinfo {author} {\bibfnamefont {J.}~\bibnamefont {Legrand}},\
  and\ \bibinfo {author} {\bibfnamefont {B.}~\bibnamefont {Macke}},\ }\bibfield
   {title} {\bibinfo {title} {Microwave echoes in inhomogeneous stark fields},\
  }\href {https://doi.org/10.1016/0009-2614(76)85080-4} {\bibfield  {journal}
  {\bibinfo  {journal} {Chemical Physics Letters}\ }\textbf {\bibinfo {volume}
  {40}},\ \bibinfo {pages} {287} (\bibinfo {year} {1976})}\BibitemShut
  {NoStop}%
\bibitem [{\citenamefont {Berman}\ \emph {et~al.}(1975)\citenamefont {Berman},
  \citenamefont {Levy},\ and\ \citenamefont
  {Brewer}}]{bermanCoherentOpticalTransient1975a}%
  \BibitemOpen
  \bibfield  {author} {\bibinfo {author} {\bibfnamefont {P.~R.}\ \bibnamefont
  {Berman}}, \bibinfo {author} {\bibfnamefont {J.~M.}\ \bibnamefont {Levy}},\
  and\ \bibinfo {author} {\bibfnamefont {R.~G.}\ \bibnamefont {Brewer}},\
  }\bibfield  {title} {\bibinfo {title} {Coherent optical transient study of
  molecular collisions: {{Theory}} and observations},\ }\href
  {https://doi.org/10.1103/PhysRevA.11.1668} {\bibfield  {journal} {\bibinfo
  {journal} {Physical Review A}\ }\textbf {\bibinfo {volume} {11}},\ \bibinfo
  {pages} {1668} (\bibinfo {year} {1975})}\BibitemShut {NoStop}%
\bibitem [{\citenamefont {Hahn}(1950)}]{hahnSpinEchoes1950}%
  \BibitemOpen
  \bibfield  {author} {\bibinfo {author} {\bibfnamefont {E.~L.}\ \bibnamefont
  {Hahn}},\ }\bibfield  {title} {\bibinfo {title} {Spin {{Echoes}}},\ }\href
  {https://doi.org/10.1103/PhysRev.80.580} {\bibfield  {journal} {\bibinfo
  {journal} {Physical Review}\ }\textbf {\bibinfo {volume} {80}},\ \bibinfo
  {pages} {580} (\bibinfo {year} {1950})}\BibitemShut {NoStop}%
\bibitem [{\citenamefont {Schneider}\ and\ \citenamefont
  {Schmiedel}(1969)}]{schneiderNegativeTimeDevelopment1969}%
  \BibitemOpen
  \bibfield  {author} {\bibinfo {author} {\bibfnamefont {H.}~\bibnamefont
  {Schneider}}\ and\ \bibinfo {author} {\bibfnamefont {H.}~\bibnamefont
  {Schmiedel}},\ }\bibfield  {title} {\bibinfo {title} {Negative time
  development of a nuclear spin system},\ }\href
  {https://doi.org/10.1016/0375-9601(69)91005-6} {\bibfield  {journal}
  {\bibinfo  {journal} {Physics Letters A}\ }\textbf {\bibinfo {volume} {30}},\
  \bibinfo {pages} {298} (\bibinfo {year} {1969})}\BibitemShut {NoStop}%
\bibitem [{\citenamefont {Rhim}\ \emph {et~al.}(1970)\citenamefont {Rhim},
  \citenamefont {Pines},\ and\ \citenamefont
  {Waugh}}]{rhimViolationSpinTemperatureHypothesis1970}%
  \BibitemOpen
  \bibfield  {author} {\bibinfo {author} {\bibfnamefont {W.-K.}\ \bibnamefont
  {Rhim}}, \bibinfo {author} {\bibfnamefont {A.}~\bibnamefont {Pines}},\ and\
  \bibinfo {author} {\bibfnamefont {J.~S.}\ \bibnamefont {Waugh}},\ }\bibfield
  {title} {\bibinfo {title} {Violation of the {{Spin-Temperature
  Hypothesis}}},\ }\href {https://doi.org/10.1103/PhysRevLett.25.218}
  {\bibfield  {journal} {\bibinfo  {journal} {Physical Review Letters}\
  }\textbf {\bibinfo {volume} {25}},\ \bibinfo {pages} {218} (\bibinfo {year}
  {1970})}\BibitemShut {NoStop}%
\bibitem [{\citenamefont {Rhim}\ \emph {et~al.}(1971)\citenamefont {Rhim},
  \citenamefont {Pines},\ and\ \citenamefont
  {Waugh}}]{rhimTimeReversalExperimentsDipolarCoupled1971a}%
  \BibitemOpen
  \bibfield  {author} {\bibinfo {author} {\bibfnamefont {W.-K.}\ \bibnamefont
  {Rhim}}, \bibinfo {author} {\bibfnamefont {A.}~\bibnamefont {Pines}},\ and\
  \bibinfo {author} {\bibfnamefont {J.~S.}\ \bibnamefont {Waugh}},\ }\bibfield
  {title} {\bibinfo {title} {Time-{{Reversal Experiments}} in {{Dipolar-Coupled
  Spin Systems}}},\ }\href {https://doi.org/10.1103/PhysRevB.3.684} {\bibfield
  {journal} {\bibinfo  {journal} {Physical Review B}\ }\textbf {\bibinfo
  {volume} {3}},\ \bibinfo {pages} {684} (\bibinfo {year} {1971})}\BibitemShut
  {NoStop}%
\bibitem [{\citenamefont {Pines}\ \emph {et~al.}(1972)\citenamefont {Pines},
  \citenamefont {Gibby},\ and\ \citenamefont
  {Waugh}}]{pinesProtonEnhancedNuclear1972}%
  \BibitemOpen
  \bibfield  {author} {\bibinfo {author} {\bibfnamefont {A.}~\bibnamefont
  {Pines}}, \bibinfo {author} {\bibfnamefont {M.~G.}\ \bibnamefont {Gibby}},\
  and\ \bibinfo {author} {\bibfnamefont {J.~S.}\ \bibnamefont {Waugh}},\
  }\bibfield  {title} {\bibinfo {title} {Proton-{{Enhanced Nuclear Induction
  Spectroscopy}}. {{A Method}} for {{High Resolution NMR}} of {{Dilute Spins}}
  in {{Solids}}},\ }\href {https://doi.org/10.1063/1.1677439} {\bibfield
  {journal} {\bibinfo  {journal} {The Journal of Chemical Physics}\ }\textbf
  {\bibinfo {volume} {56}},\ \bibinfo {pages} {1776} (\bibinfo {year}
  {1972})}\BibitemShut {NoStop}%
\bibitem [{\citenamefont {Hartmann}\ and\ \citenamefont
  {Hahn}(1962)}]{hartmannNuclearDoubleResonance1962}%
  \BibitemOpen
  \bibfield  {author} {\bibinfo {author} {\bibfnamefont {S.~R.}\ \bibnamefont
  {Hartmann}}\ and\ \bibinfo {author} {\bibfnamefont {E.~L.}\ \bibnamefont
  {Hahn}},\ }\bibfield  {title} {\bibinfo {title} {Nuclear {{Double Resonance}}
  in the {{Rotating Frame}}},\ }\href
  {https://doi.org/10.1103/PhysRev.128.2042} {\bibfield  {journal} {\bibinfo
  {journal} {Physical Review}\ }\textbf {\bibinfo {volume} {128}},\ \bibinfo
  {pages} {2042} (\bibinfo {year} {1962})}\BibitemShut {NoStop}%
\bibitem [{\citenamefont {Ernst}\ \emph {et~al.}(1998)\citenamefont {Ernst},
  \citenamefont {Meier}, \citenamefont {Tomaselli},\ and\ \citenamefont
  {Pines}}]{ernstTimereversalCrosspolarizationNuclear1998}%
  \BibitemOpen
  \bibfield  {author} {\bibinfo {author} {\bibfnamefont {M.}~\bibnamefont
  {Ernst}}, \bibinfo {author} {\bibfnamefont {B.~H.}\ \bibnamefont {Meier}},
  \bibinfo {author} {\bibfnamefont {M.}~\bibnamefont {Tomaselli}},\ and\
  \bibinfo {author} {\bibfnamefont {A.}~\bibnamefont {Pines}},\ }\bibfield
  {title} {\bibinfo {title} {Time-reversal of cross-polarization in nuclear
  magnetic resonance},\ }\href {https://doi.org/10.1063/1.476435} {\bibfield
  {journal} {\bibinfo  {journal} {The Journal of Chemical Physics}\ }\textbf
  {\bibinfo {volume} {108}},\ \bibinfo {pages} {9611} (\bibinfo {year}
  {1998})}\BibitemShut {NoStop}%
\bibitem [{\citenamefont {Abragam}\ and\ \citenamefont
  {Goldman}(1978)}]{abragamPrinciplesDynamicNuclear1978}%
  \BibitemOpen
  \bibfield  {author} {\bibinfo {author} {\bibfnamefont {A.}~\bibnamefont
  {Abragam}}\ and\ \bibinfo {author} {\bibfnamefont {M.}~\bibnamefont
  {Goldman}},\ }\bibfield  {title} {\bibinfo {title} {Principles of dynamic
  nuclear polarisation},\ }\href {https://doi.org/10.1088/0034-4885/41/3/002}
  {\bibfield  {journal} {\bibinfo  {journal} {Reports on Progress in Physics}\
  }\textbf {\bibinfo {volume} {41}},\ \bibinfo {pages} {395} (\bibinfo {year}
  {1978})}\BibitemShut {NoStop}%
\bibitem [{\citenamefont {Maly}\ \emph {et~al.}(2008)\citenamefont {Maly},
  \citenamefont {Debelouchina}, \citenamefont {Bajaj}, \citenamefont {Hu},
  \citenamefont {Joo}, \citenamefont {{Mak--Jurkauskas}}, \citenamefont
  {Sirigiri}, \citenamefont {{van der Wel}}, \citenamefont {Herzfeld},
  \citenamefont {Temkin},\ and\ \citenamefont
  {Griffin}}]{malyDynamicNuclearPolarization2008}%
  \BibitemOpen
  \bibfield  {author} {\bibinfo {author} {\bibfnamefont {T.}~\bibnamefont
  {Maly}}, \bibinfo {author} {\bibfnamefont {G.~T.}\ \bibnamefont
  {Debelouchina}}, \bibinfo {author} {\bibfnamefont {V.~S.}\ \bibnamefont
  {Bajaj}}, \bibinfo {author} {\bibfnamefont {K.-N.}\ \bibnamefont {Hu}},
  \bibinfo {author} {\bibfnamefont {C.-G.}\ \bibnamefont {Joo}}, \bibinfo
  {author} {\bibfnamefont {M.~L.}\ \bibnamefont {{Mak--Jurkauskas}}}, \bibinfo
  {author} {\bibfnamefont {J.~R.}\ \bibnamefont {Sirigiri}}, \bibinfo {author}
  {\bibfnamefont {P.~C.~A.}\ \bibnamefont {{van der Wel}}}, \bibinfo {author}
  {\bibfnamefont {J.}~\bibnamefont {Herzfeld}}, \bibinfo {author}
  {\bibfnamefont {R.~J.}\ \bibnamefont {Temkin}},\ and\ \bibinfo {author}
  {\bibfnamefont {R.~G.}\ \bibnamefont {Griffin}},\ }\bibfield  {title}
  {\bibinfo {title} {Dynamic nuclear polarization at high magnetic fields},\
  }\href {https://doi.org/10.1063/1.2833582} {\bibfield  {journal} {\bibinfo
  {journal} {The Journal of Chemical Physics}\ }\textbf {\bibinfo {volume}
  {128}},\ \bibinfo {pages} {052211} (\bibinfo {year} {2008})}\BibitemShut
  {NoStop}%
\bibitem [{\citenamefont {Lilly~Thankamony}\ \emph {et~al.}(2017)\citenamefont
  {Lilly~Thankamony}, \citenamefont {Wittmann}, \citenamefont {Kaushik},\ and\
  \citenamefont {Corzilius}}]{lillythankamonyDynamicNuclearPolarization2017}%
  \BibitemOpen
  \bibfield  {author} {\bibinfo {author} {\bibfnamefont {A.~S.}\ \bibnamefont
  {Lilly~Thankamony}}, \bibinfo {author} {\bibfnamefont {J.~J.}\ \bibnamefont
  {Wittmann}}, \bibinfo {author} {\bibfnamefont {M.}~\bibnamefont {Kaushik}},\
  and\ \bibinfo {author} {\bibfnamefont {B.}~\bibnamefont {Corzilius}},\
  }\bibfield  {title} {\bibinfo {title} {Dynamic nuclear polarization for
  sensitivity enhancement in modern solid-state {{NMR}}},\ }\href
  {https://doi.org/10.1016/j.pnmrs.2017.06.002} {\bibfield  {journal} {\bibinfo
   {journal} {Progress in Nuclear Magnetic Resonance Spectroscopy}\ }\textbf
  {\bibinfo {volume} {102--103}},\ \bibinfo {pages} {120} (\bibinfo {year}
  {2017})}\BibitemShut {NoStop}%
\bibitem [{\citenamefont {Brunner}\ \emph {et~al.}(1987)\citenamefont
  {Brunner}, \citenamefont {Fritsch},\ and\ \citenamefont
  {Hausser}}]{brunnerCrossPolarizationElectron1987}%
  \BibitemOpen
  \bibfield  {author} {\bibinfo {author} {\bibfnamefont {H.}~\bibnamefont
  {Brunner}}, \bibinfo {author} {\bibfnamefont {R.~H.}\ \bibnamefont
  {Fritsch}},\ and\ \bibinfo {author} {\bibfnamefont {K.~H.}\ \bibnamefont
  {Hausser}},\ }\bibfield  {title} {\bibinfo {title} {Cross {{Polarization}} in
  {{Electron Nuclear Double Resonance}} by {{Satisfying}} the {{Hartmann-Hahn
  Condition}}},\ }\href {https://doi.org/10.1515/zna-1987-1217} {\bibfield
  {journal} {\bibinfo  {journal} {Zeitschrift f{\"u}r Naturforschung A}\
  }\textbf {\bibinfo {volume} {42}},\ \bibinfo {pages} {1456} (\bibinfo {year}
  {1987})}\BibitemShut {NoStop}%
\bibitem [{\citenamefont {Henstra}\ \emph {et~al.}(1988)\citenamefont
  {Henstra}, \citenamefont {Dirksen}, \citenamefont {Schmidt},\ and\
  \citenamefont {Wenckebach}}]{henstraNuclearSpinOrientation1988}%
  \BibitemOpen
  \bibfield  {author} {\bibinfo {author} {\bibfnamefont {A.}~\bibnamefont
  {Henstra}}, \bibinfo {author} {\bibfnamefont {P.}~\bibnamefont {Dirksen}},
  \bibinfo {author} {\bibfnamefont {J.}~\bibnamefont {Schmidt}},\ and\ \bibinfo
  {author} {\bibfnamefont {W.}~\bibnamefont {Wenckebach}},\ }\bibfield  {title}
  {\bibinfo {title} {Nuclear spin orientation via electron spin locking
  ({{NOVEL}})},\ }\href {https://doi.org/10.1016/0022-2364(88)90190-4}
  {\bibfield  {journal} {\bibinfo  {journal} {Journal of Magnetic Resonance
  (1969)}\ }\textbf {\bibinfo {volume} {77}},\ \bibinfo {pages} {389} (\bibinfo
  {year} {1988})}\BibitemShut {NoStop}%
\bibitem [{\citenamefont {Schwartz}\ \emph {et~al.}(2018)\citenamefont
  {Schwartz}, \citenamefont {Scheuer}, \citenamefont {Tratzmiller},
  \citenamefont {M{\"u}ller}, \citenamefont {Chen}, \citenamefont {Dhand},
  \citenamefont {Wang}, \citenamefont {M{\"u}ller}, \citenamefont {Naydenov},
  \citenamefont {Jelezko},\ and\ \citenamefont
  {Plenio}}]{schwartzRobustOpticalPolarization2018}%
  \BibitemOpen
  \bibfield  {author} {\bibinfo {author} {\bibfnamefont {I.}~\bibnamefont
  {Schwartz}}, \bibinfo {author} {\bibfnamefont {J.}~\bibnamefont {Scheuer}},
  \bibinfo {author} {\bibfnamefont {B.}~\bibnamefont {Tratzmiller}}, \bibinfo
  {author} {\bibfnamefont {S.}~\bibnamefont {M{\"u}ller}}, \bibinfo {author}
  {\bibfnamefont {Q.}~\bibnamefont {Chen}}, \bibinfo {author} {\bibfnamefont
  {I.}~\bibnamefont {Dhand}}, \bibinfo {author} {\bibfnamefont {Z.-Y.}\
  \bibnamefont {Wang}}, \bibinfo {author} {\bibfnamefont {C.}~\bibnamefont
  {M{\"u}ller}}, \bibinfo {author} {\bibfnamefont {B.}~\bibnamefont
  {Naydenov}}, \bibinfo {author} {\bibfnamefont {F.}~\bibnamefont {Jelezko}},\
  and\ \bibinfo {author} {\bibfnamefont {M.~B.}\ \bibnamefont {Plenio}},\
  }\bibfield  {title} {\bibinfo {title} {Robust optical polarization of nuclear
  spin baths using {{Hamiltonian}} engineering of nitrogen-vacancy center
  quantum dynamics},\ }\href {https://doi.org/10.1126/sciadv.aat8978}
  {\bibfield  {journal} {\bibinfo  {journal} {Science Advances}\ }\textbf
  {\bibinfo {volume} {4}},\ \bibinfo {pages} {eaat8978} (\bibinfo {year}
  {2018})}\BibitemShut {NoStop}%
\bibitem [{\citenamefont {Tan}\ \emph {et~al.}(2019)\citenamefont {Tan},
  \citenamefont {Yang}, \citenamefont {Weber}, \citenamefont {Mathies},\ and\
  \citenamefont {Griffin}}]{tanTimeoptimizedPulsedDynamic2019}%
  \BibitemOpen
  \bibfield  {author} {\bibinfo {author} {\bibfnamefont {K.~O.}\ \bibnamefont
  {Tan}}, \bibinfo {author} {\bibfnamefont {C.}~\bibnamefont {Yang}}, \bibinfo
  {author} {\bibfnamefont {R.~T.}\ \bibnamefont {Weber}}, \bibinfo {author}
  {\bibfnamefont {G.}~\bibnamefont {Mathies}},\ and\ \bibinfo {author}
  {\bibfnamefont {R.~G.}\ \bibnamefont {Griffin}},\ }\bibfield  {title}
  {\bibinfo {title} {Time-optimized pulsed dynamic nuclear polarization},\
  }\href {https://doi.org/10.1126/sciadv.aav6909} {\bibfield  {journal}
  {\bibinfo  {journal} {Science Advances}\ }\textbf {\bibinfo {volume} {5}},\
  \bibinfo {pages} {eaav6909} (\bibinfo {year} {2019})}\BibitemShut {NoStop}%
\bibitem [{\citenamefont {Redrouthu}\ and\ \citenamefont
  {Mathies}(2022)}]{redrouthuEfficientPulsedDynamic2022}%
  \BibitemOpen
  \bibfield  {author} {\bibinfo {author} {\bibfnamefont {V.~S.}\ \bibnamefont
  {Redrouthu}}\ and\ \bibinfo {author} {\bibfnamefont {G.}~\bibnamefont
  {Mathies}},\ }\bibfield  {title} {\bibinfo {title} {Efficient {{Pulsed
  Dynamic Nuclear Polarization}} with the {{X-Inverse-X Sequence}}},\ }\href
  {https://doi.org/10.1021/jacs.1c09900} {\bibfield  {journal} {\bibinfo
  {journal} {Journal of the American Chemical Society}\ }\textbf {\bibinfo
  {volume} {144}},\ \bibinfo {pages} {1513} (\bibinfo {year}
  {2022})}\BibitemShut {NoStop}%
\bibitem [{\citenamefont {Wili}\ \emph {et~al.}(2022)\citenamefont {Wili},
  \citenamefont {Nielsen}, \citenamefont {V{\"o}lker}, \citenamefont
  {Schreder}, \citenamefont {Nielsen}, \citenamefont {Jeschke},\ and\
  \citenamefont {Tan}}]{wiliDesigningBroadbandPulsed2022}%
  \BibitemOpen
  \bibfield  {author} {\bibinfo {author} {\bibfnamefont {N.}~\bibnamefont
  {Wili}}, \bibinfo {author} {\bibfnamefont {A.~B.}\ \bibnamefont {Nielsen}},
  \bibinfo {author} {\bibfnamefont {L.~A.}\ \bibnamefont {V{\"o}lker}},
  \bibinfo {author} {\bibfnamefont {L.}~\bibnamefont {Schreder}}, \bibinfo
  {author} {\bibfnamefont {N.~C.}\ \bibnamefont {Nielsen}}, \bibinfo {author}
  {\bibfnamefont {G.}~\bibnamefont {Jeschke}},\ and\ \bibinfo {author}
  {\bibfnamefont {K.~O.}\ \bibnamefont {Tan}},\ }\bibfield  {title} {\bibinfo
  {title} {Designing broadband pulsed dynamic nuclear polarization sequences in
  static solids},\ }\href {https://doi.org/10.1126/sciadv.abq0536} {\bibfield
  {journal} {\bibinfo  {journal} {Science Advances}\ }\textbf {\bibinfo
  {volume} {8}},\ \bibinfo {pages} {eabq0536} (\bibinfo {year}
  {2022})}\BibitemShut {NoStop}%
\bibitem [{\citenamefont {Redrouthu}\ \emph {et~al.}(2023)\citenamefont
  {Redrouthu}, \citenamefont {{Vinod-Kumar}},\ and\ \citenamefont
  {Mathies}}]{redrouthuDynamicNuclearPolarization2023}%
  \BibitemOpen
  \bibfield  {author} {\bibinfo {author} {\bibfnamefont {V.~S.}\ \bibnamefont
  {Redrouthu}}, \bibinfo {author} {\bibfnamefont {S.}~\bibnamefont
  {{Vinod-Kumar}}},\ and\ \bibinfo {author} {\bibfnamefont {G.}~\bibnamefont
  {Mathies}},\ }\bibfield  {title} {\bibinfo {title} {Dynamic nuclear
  polarization by two-pulse phase modulation},\ }\href
  {https://doi.org/10.1063/5.0153053} {\bibfield  {journal} {\bibinfo
  {journal} {The Journal of Chemical Physics}\ }\textbf {\bibinfo {volume}
  {159}},\ \bibinfo {pages} {014201} (\bibinfo {year} {2023})}\BibitemShut
  {NoStop}%
\bibitem [{\citenamefont {Shankar}\ \emph {et~al.}(2017)\citenamefont
  {Shankar}, \citenamefont {Ernst}, \citenamefont {Madhu}, \citenamefont
  {Vosegaard}, \citenamefont {Nielsen},\ and\ \citenamefont
  {Nielsen}}]{shankarGeneralTheoreticalDescription2017}%
  \BibitemOpen
  \bibfield  {author} {\bibinfo {author} {\bibfnamefont {R.}~\bibnamefont
  {Shankar}}, \bibinfo {author} {\bibfnamefont {M.}~\bibnamefont {Ernst}},
  \bibinfo {author} {\bibfnamefont {P.~K.}\ \bibnamefont {Madhu}}, \bibinfo
  {author} {\bibfnamefont {T.}~\bibnamefont {Vosegaard}}, \bibinfo {author}
  {\bibfnamefont {N.~C.}\ \bibnamefont {Nielsen}},\ and\ \bibinfo {author}
  {\bibfnamefont {A.~B.}\ \bibnamefont {Nielsen}},\ }\bibfield  {title}
  {\bibinfo {title} {A general theoretical description of the influence of
  isotropic chemical shift in dipolar recoupling experiments for solid-state
  {{NMR}}},\ }\href {https://doi.org/10.1063/1.4979123} {\bibfield  {journal}
  {\bibinfo  {journal} {The Journal of Chemical Physics}\ }\textbf {\bibinfo
  {volume} {146}},\ \bibinfo {pages} {134105} (\bibinfo {year}
  {2017})}\BibitemShut {NoStop}%
\bibitem [{\citenamefont {Nielsen}\ \emph {et~al.}(2019)\citenamefont
  {Nielsen}, \citenamefont {Hansen}, \citenamefont {Andersen},\ and\
  \citenamefont {Vosegaard}}]{nielsenSinglespinVectorAnalysis2019}%
  \BibitemOpen
  \bibfield  {author} {\bibinfo {author} {\bibfnamefont {A.~B.}\ \bibnamefont
  {Nielsen}}, \bibinfo {author} {\bibfnamefont {M.~R.}\ \bibnamefont {Hansen}},
  \bibinfo {author} {\bibfnamefont {J.~E.}\ \bibnamefont {Andersen}},\ and\
  \bibinfo {author} {\bibfnamefont {T.}~\bibnamefont {Vosegaard}},\ }\bibfield
  {title} {\bibinfo {title} {Single-spin vector analysis of strongly coupled
  nuclei in {{TOCSY NMR}} experiments},\ }\href
  {https://doi.org/10.1063/1.5123046} {\bibfield  {journal} {\bibinfo
  {journal} {The Journal of Chemical Physics}\ }\textbf {\bibinfo {volume}
  {151}},\ \bibinfo {pages} {134117} (\bibinfo {year} {2019})}\BibitemShut
  {NoStop}%
\bibitem [{\citenamefont {Nielsen}\ and\ \citenamefont
  {Nielsen}(2022)}]{nielsenAccurateAnalysisPerspectives2022}%
  \BibitemOpen
  \bibfield  {author} {\bibinfo {author} {\bibfnamefont {A.~B.}\ \bibnamefont
  {Nielsen}}\ and\ \bibinfo {author} {\bibfnamefont {N.~C.}\ \bibnamefont
  {Nielsen}},\ }\bibfield  {title} {\bibinfo {title} {Accurate analysis and
  perspectives for systematic design of magnetic resonance experiments using
  single-spin vector and exact effective {{Hamiltonian}} theory},\ }\href
  {https://doi.org/10.1016/j.jmro.2022.100064} {\bibfield  {journal} {\bibinfo
  {journal} {Journal of Magnetic Resonance Open}\ }\textbf {\bibinfo {volume}
  {12--13}},\ \bibinfo {pages} {100064} (\bibinfo {year} {2022})}\BibitemShut
  {NoStop}%
\bibitem [{Note1()}]{Note1}%
  \BibitemOpen
  \bibinfo {note} {Another special case is $|\omega _\protect \text
  {eff}^{(S)}|\approx |\omega _\protect \text {eff}^{(I)}|\approx \omega
  _\protect \text {m}/2$. In this case there are additional time-independent
  terms in the interaction frame that have to be considered. However, in this
  case one could look at two periods of the repeating element, which would
  again lead to $|\omega _\protect \text {eff}^{(S)}|\approx |\omega _\protect
  \text {eff}^{(I)}|\approx 0$.}\BibitemShut {Stop}%
\bibitem [{\citenamefont {Doll}\ and\ \citenamefont
  {Jeschke}(2017)}]{dollWidebandFrequencysweptExcitation2017}%
  \BibitemOpen
  \bibfield  {author} {\bibinfo {author} {\bibfnamefont {A.}~\bibnamefont
  {Doll}}\ and\ \bibinfo {author} {\bibfnamefont {G.}~\bibnamefont {Jeschke}},\
  }\bibfield  {title} {\bibinfo {title} {Wideband frequency-swept excitation in
  pulsed {{EPR}} spectroscopy},\ }\href
  {https://doi.org/10.1016/j.jmr.2017.01.004} {\bibfield  {journal} {\bibinfo
  {journal} {Journal of Magnetic Resonance}\ }\textbf {\bibinfo {volume}
  {280}},\ \bibinfo {pages} {46} (\bibinfo {year} {2017})}\BibitemShut
  {NoStop}%
\bibitem [{\citenamefont {Mathies}\ \emph {et~al.}(2016)\citenamefont
  {Mathies}, \citenamefont {Jain}, \citenamefont {Reese},\ and\ \citenamefont
  {Griffin}}]{mathiesPulsedDynamicNuclear2016}%
  \BibitemOpen
  \bibfield  {author} {\bibinfo {author} {\bibfnamefont {G.}~\bibnamefont
  {Mathies}}, \bibinfo {author} {\bibfnamefont {S.}~\bibnamefont {Jain}},
  \bibinfo {author} {\bibfnamefont {M.}~\bibnamefont {Reese}},\ and\ \bibinfo
  {author} {\bibfnamefont {R.~G.}\ \bibnamefont {Griffin}},\ }\bibfield
  {title} {\bibinfo {title} {Pulsed {{Dynamic Nuclear Polarization}} with
  {{Trityl Radicals}}},\ }\href {https://doi.org/10.1021/acs.jpclett.5b02720}
  {\bibfield  {journal} {\bibinfo  {journal} {The Journal of Physical Chemistry
  Letters}\ }\textbf {\bibinfo {volume} {7}},\ \bibinfo {pages} {111} (\bibinfo
  {year} {2016})}\BibitemShut {NoStop}%
\bibitem [{\citenamefont {Degen}\ \emph {et~al.}(2017)\citenamefont {Degen},
  \citenamefont {Reinhard},\ and\ \citenamefont
  {Cappellaro}}]{degenQuantumSensing2017}%
  \BibitemOpen
  \bibfield  {author} {\bibinfo {author} {\bibfnamefont {C.~L.}\ \bibnamefont
  {Degen}}, \bibinfo {author} {\bibfnamefont {F.}~\bibnamefont {Reinhard}},\
  and\ \bibinfo {author} {\bibfnamefont {P.}~\bibnamefont {Cappellaro}},\
  }\bibfield  {title} {\bibinfo {title} {Quantum sensing},\ }\href
  {https://doi.org/10.1103/RevModPhys.89.035002} {\bibfield  {journal}
  {\bibinfo  {journal} {Reviews of Modern Physics}\ }\textbf {\bibinfo {volume}
  {89}},\ \bibinfo {pages} {035002} (\bibinfo {year} {2017})}\BibitemShut
  {NoStop}%
\bibitem [{\citenamefont {Gruber}\ \emph {et~al.}(1997)\citenamefont {Gruber},
  \citenamefont {Dr{\"a}benstedt}, \citenamefont {Tietz}, \citenamefont
  {Fleury}, \citenamefont {Wrachtrup},\ and\ \citenamefont {von
  Borczyskowski}}]{gruberScanningConfocalOptical1997}%
  \BibitemOpen
  \bibfield  {author} {\bibinfo {author} {\bibfnamefont {A.}~\bibnamefont
  {Gruber}}, \bibinfo {author} {\bibfnamefont {A.}~\bibnamefont
  {Dr{\"a}benstedt}}, \bibinfo {author} {\bibfnamefont {C.}~\bibnamefont
  {Tietz}}, \bibinfo {author} {\bibfnamefont {L.}~\bibnamefont {Fleury}},
  \bibinfo {author} {\bibfnamefont {J.}~\bibnamefont {Wrachtrup}},\ and\
  \bibinfo {author} {\bibfnamefont {C.}~\bibnamefont {von Borczyskowski}},\
  }\bibfield  {title} {\bibinfo {title} {Scanning {{Confocal Optical
  Microscopy}} and {{Magnetic Resonance}} on {{Single Defect Centers}}},\
  }\href {https://doi.org/10.1126/science.276.5321.2012} {\bibfield  {journal}
  {\bibinfo  {journal} {Science}\ }\textbf {\bibinfo {volume} {276}},\ \bibinfo
  {pages} {2012} (\bibinfo {year} {1997})}\BibitemShut {NoStop}%
\end{thebibliography}%

\end{document}